\documentclass[a4paper,12pt]{article}
\usepackage{amsmath,amssymb}
\usepackage{bm}
\usepackage{color}
\usepackage{cite}
\usepackage{mathtools}

\usepackage{multirow}

\usepackage{here}
\usepackage {cancel}

\begin{document}

\begin{titlepage}


\vskip 1.35cm
\begin{center}
{\Large \bf   Axionless Solution to the Strong CP Problem \\
    -- two-zeros textures of the quark and lepton \\  mass matrices and neutrino CP violation -- }
    \end{center}
\vskip 1.2cm
\begin{center}
{\large Morimitsu Tanimoto} $^{a}$
	\footnote{email: morimitsutanimoto@yahoo.co.jp}
	\hskip 0.3 cm 
 {\large and Tsutomu T. Yanagida} $^{b,\,c}$
 \footnote{email: tsutomu.tyanagida@sjtu.edu.cn 
 }
\end{center}	
\centerline{
		\begin{minipage}{\linewidth}
			\begin{center}
					$^a${\it \normalsize
					Department of Physics, Niigata University, Niigata 950-2181, Japan }
				\\*[10pt]
			\end{center}
      \begin{center}
					$^b${\it \normalsize
					Kavli Institute for the Physics and Mathematics of Universe (WPI),
     \\*[10pt] University of Tokyo, Kashiwa 277-8583, Japan }
				\\*[10pt]
			\end{center}
    \begin{center}
					$^c${\it \normalsize
	Tsung-Dao Lee Institute and School of Physics and Astronomy,
     \\*[10pt] 
      Shanghai Jiao Tong University, China}
				\\*[10pt]
			\end{center}
	\end{minipage}
    }
    \vskip 0.7 cm
 \begin{center}
\today
 \end{center}
\vskip 0.8 cm
	\begin{minipage}{0.94\linewidth}
\abstract{
CP invariance is a very attractive solution to the strong CP problem in QCD. This solution requires the vanishing ${\rm arg}\,[{\rm det}\, M_d\, {\rm det} M_u]$, where the $M_d$ and $M_u$ are the mass matrices for the down- and up-type quarks. It happens if we have several zeros in the quark mass matrices. We proceed a systematic construction, in this paper, of two zeros textures for the down-type quark mass matrix while the mass matrix for the up-type quarks is always diagonal. We find only three types of the mass matrices can explain the observed CKM matrix, the masses of the quarks and the charged leptons and the small enough vacuum angle $\theta < 10^{-10}$. We extend the mass construction to the neutrino sector and derive predictions on the CP violating parameter $\delta_{CP}$ in the neutrino oscillation and the mass parameter $m_{\beta\beta}$. It is extremely remarkable that the normal (NH) and inverted (IH) hierarchies in the neutrino masses are equally possible in the case where we introduce only two right-handed neutrinos $N$s. Furthermore, we have a strict prediction on the $\delta_{CP} \simeq 200^\circ$ or $250^\circ$ in the NH case. If it is the case we can naturally explain the positive sign of the baryon asymmetry in the present universe.
}
\end{minipage}

\end{titlepage}

\setcounter{page}{2}

\section{Introduction}

The CP or equivalently time-reversal invariance is a very attractive solution to the strong CP problem in QCD. However, it was argued that the vacuum angle $\theta$ is shifted from 0 by the diagonalization of quark mass matrices if ${\rm arg}\,[{\rm det}\, M_d\, {\rm det} M_u]$ is not vanishing, where the $M_d$ and $M_u$ are the mass matrices for the down- and up-type quarks. Recently, it was pointed \cite{Liang:2024wbb} that we can naturally build three-zeros textures for the quark mass matrices whose determinants are always real \cite{Tanimoto:2016rqy}. The model is based on the six-dimensional spacetime with the $\bm {T^2/Z_3}$ orbifold compactification.

In the previous paper \cite{Tanimoto:2024nwm} we  extended the above LOY model \cite{Liang:2024wbb} to the lepton sector to predict the CP violating phase $\delta_{CP}$ in the neutrino oscillation. And the key mass parameter  $m_{\beta\beta}$ for double $\beta$ decay amplitudes of nuclei was also predicted. 

However,  even two zeros textures can provide a solution to the strong CP problem as pointed out in \cite{Liang:2024wbb} . We show, in this paper, a systematic construction of the two zeros textures for the quark and lepton mass matrices in the same setup with the  $\bm {T^2/Z_3}$ orbifold compactification of the extra two dimensions. We show that only three types of the mass matrices can explain the observed CKM matrix, the masses of the quarks and the charged leptons and the small enough vacuum angle $\theta < 10^{-10}$. 

We extend the mass construction to the neutrino sector and  discuss the predictions on the CP violation parameter $\delta_{CP}$ measurable in the neutrino oscillation experiments and the key parameter $m_{\beta\beta}$  as done in \cite{Tanimoto:2024nwm}, introducing three right-handed neutrinos $N_i~(i=1,2,3)$. We find the normal hierarchy (NH) is naturally realized as expected.  

We stress here, however,  that the inverted hierarchy (IH) in the neutrino masses is even naturally built in the limit of the Majorana mass $M_3=\infty$ where the $M_3$ is the mass of the third right-handed neutrino, against the long standing folktale. The $\delta_{CP}$ is predicted in a broad region. On the other hand we realize naturally the normal hierarchy (NH) in the limit of Majorana mass for the first family right-handed neutrino being infinity $M_1=\infty$. In this case we have a strict prediction on the $\delta_{CP} \simeq 200^\circ$ or $250^\circ$. It is also very interesting that we can naturally explain the correct (positive) sign of the baryon asymmetry in the present universe.

{
We consider, throughout this paper, the CP invariance at the fundamental level, which is, however, spontaneously broken down at some intermediate scale such as $10^{12}$ GeV, for example. The spontaneous CP violation is necessary to explain the observed CP violation in the CKM matrix and to generate the baryon asymmetry in the present universe
\footnote{String theory has 
CP symmetry \cite{Green:1987mn} before the compactification.
One possible scenario for 
spontaneous CP violation is moduli stabilization through 
compactifications. However, a simple moduli stabilization leads to CP 
invariant vacua. In particular, if the theory has more symmetries, e.g. 
modular symmetry in addition to the CP symmetry, those symmetries tend 
to protect the CP symmetry from its violation.
(See e.g. Refs. \cite{Kobayashi:2019uyt,Kobayashi:2020uaj,Ishiguro:2020nuf,Novichkov:2022wvg,Knapp-Perez:2023nty}.) 
Therefore, we consider a scenario that the CP is spontaneously broken 
down in the framework of four-dimensional effective field theory, 
in this paper.}
. We construct the diagonal texture for the up-type quark mass matrix and the two zero textures for the down-type quark mass matrix at the very high energy scale, which satisfy ${\rm arg}\,[{\rm det}\, M_d\, {\rm det} M_u]$=0 so that the physical vacuum angle ${\bar \theta}$ =0. It is crucial that the physical vacuum angle ${\bar \theta}$ remains sufficiently small at low energies once the above condition is satisfied at some high energy scale, as proved in \cite{Ellis:1978hq}.
}


\section{Construction of two-zeros textures for the quark and lepton mass matrices}

The LOY construction \cite{Liang:2024wbb} of three-zeros textures for quark mass matrices is based on the orbifold compactification $\bm{T^2/Z_3}$ of the six dimensional space-time and a $\bm{Z_2}$ flavor symmetry. We adopt the same setting assumption in this paper. In addition we assume the $SU(5)_{GUT}$ representations for all the fermions. But, we gauge only the standard model (SM) group, $SU(3)\times SU(2)\times U(1)_Y$, that is a subgroup of the $SU(5)_{GUT}$.

There are three fixed points, I, I\hskip -0.5mm I and
I\hskip -0.5mm I\hskip -0.5mm I, in the extra two dimensions of $\bm{T^2/Z_3}$. We put each ${\bm{10_i~(i=1,2,3)}}$ on the each fixed point, separately. We assume all $\mathbf{5^*_i~(i=1,2,3)}$ and the SM Higgs $\mathbf{H}$ are living in the two dimensional bulk. We assume the Higgs $\mathbf{H}$ belongs to a $\mathbf{5}$ representation of the $SU(5)_{GUT}$ and $\mathbf{10_i 10_j H}$ Yukawa coupling at the fixed points. We realize a diagonal mass matrix for the up-type quarks by taking the size of the $\mathbf{T^2/Z_3}$ orbifold dimensions sufficiently large \cite{Liang:2024wbb}. Notice that all masses for the up-type quarks are real, since all Yukawa coupling constants are real because of our basic assumption of the CP invariance at high energies. The vacuum-expectation value (vev) of the Higgs, $\mathbf{<H>}$, can be always taken real by using the $U(1)_Y$ hypercharge gauge transformation. 

We now impose an anomaly free discrete $\mathbf{Z_2}$ flavor symmetry to have several zeros in the mass matrix for the down-type quarks. There are four choices for the $\mathbf{Z_2}$ charges for the $\mathbf{10}$s, that is, three 
$\mathbf{+}$ s and zero $\bm {-}$; two $\bm{+}$s and one $\bm{-}$; one $\bm{+}$ and two $\bm{-}$s; zero $\bm{+}$ and three $\bm{-}$s. 
However, the first and fourth choices are unable to reproduce the observed CKM matrix with a real determinant of the down-type quark mass matrix, and the second and third choices are essentially equivalent. Thus, we have only one choice for the $\mathbf{Z_2}$ charge. We fix it as two $\bm{+}$s and one ${\bm{-}}$. Then, the charges of the $\bm{5^*}$ are determined as two ${\bm +}$s and one ${\bm -}$ so that the discrete $\mathbf{Z_2}$ is free from anomaly
\footnote{ The choice of three $\bm{-}$s is also anomaly free, but it can not reproduce the correct CKM matrix with a real determinant of the down-type quark mass matrix.}.
We take the SM Higgs to be even($\bm{+}$) for the $\mathbf{Z_2}$.

We have three generations for the $\bm{10}$s, that is, $\bm{10_1}$, $\bm{10_2}$ and $\bm{10_3}$. Thus, we have three assignments of the $\bm{Z_2}$ charges as $\bm{10_{1,2,3}}= (\bm{-},\bm{+},\bm{+}); (\bm{+},\bm{-},\bm{+}); (\bm{+},\bm{+},\bm{-})$ \footnote{We name $\bm{10_{1,2,3}}$ in order of mass.}. And we have three charge assignment for the $\bm{5^*}$s  as $\bm{5^*_{1,2,3}}= (\bm{-},\bm{+},\bm{+}); (\bm{+},\bm{-},\bm{+}); (\bm{+},\bm{+},\bm{-}) $, too. Therefore, we have $3\times 3=9$ assignments for the $\mathbf{Z_2}$ charge in total. However, it turns out that only the case of $\bm{10_{1,2,3}}=(\bm{-},\bm{+},\bm{+}) $ can reproduce correctly the observed CKM matrix and charged lepton mass ratios. Thus, we have only three type of the charge assignments called $\bm{A_i}$ type: 
\begin{align}
&\bm{A_1; 10_{1,2,3}}=(\bm{-},\bm{+},\bm{+}), ~~~\bm{5^*_{1,2,3}}= (\bm{+},\bm{-},\bm{+});\nonumber\\
&\bm{A_2; 10_{1,2,3}= (\bm{-},\bm{+},\bm{+})}, ~~\bm{5^*_{1,2,3}=(\bm{+},\bm{+},\bm{-})} ;\nonumber\\
&\bm{A_3; 10_{1,2,3}=(\bm{-},\bm{+},\bm{+})} , ~~\bm{5^*_{1,2,3}=(\bm{-},\bm{+},\bm{+})}.
\label{Ai-type}
\end{align}

We show, in Table 1,  $\bm{Z_2}$  charges and  locations  in the extra two dimensions for each particles for the case of $\bm{A_1}$.
For the cases of $\bm{A_2}$ and $\bm{A_3}$, 
the only  $\bm{Z_2}$  charges of  $\bm{5^*_{1,2,3}}$ are replaced as shown in Eq.\,\eqref{list}.

\begin{table}[H]
\begin{center}
\begin{tabular}{|c||c|c|c|c|c|c|c|c|c|c|c|c|}
\hline 
	\rule[14pt]{0pt}{3pt}  
 Case $\bm{A_1}$&$\bf 10_1$ &$\bf 10_2$ &$\bf 10_3$ & $\bf 5_1^*$ &$\bf 5_2^*$ & $\bf 5_3^*$ & $N_{1}$ &
$N_{2}$& $N_{3}$& $H$ &  $\eta$  &$\eta'$\\
\hline 
	\rule[14pt]{0pt}{3pt}  
$\bm{Z_2}$ &  - & + & + & + & - & + & - & + & + & + & - & -\\
\hline 
	\rule[14pt]{0pt}{3pt}  
location&  I & I\hskip -0.5mm I &  I\hskip -0.5mm I\hskip -0.5mm I 
& bulk & bulk & bulk & I & I\hskip -0.5mm I &  I\hskip -0.5mm I\hskip -0.5mm I &  bulk &    I\hskip -0.5mm I &  I\hskip -0.5mm I\hskip -0.5mm I \\
\hline
\end{tabular}
\caption{List of $\bm{Z_2}$  charges and  locations  in the extra two dimensions
(fixed points I, I\hskip -0.5mm I, I\hskip -0.5mm I\hskip -0.5mm I  and bulk) for each particles in the case of $\bm{A_1}$,
where $N_i$, $\eta$ and $\eta'$ are introduced in the following subsections.}
\label{list}
\end{center}
\end{table}


\subsection{The down-type quark mass matrices}

We explain the construction of the mass matrices for the down-type quarks mostly in the case of the type $\bm{A_1}$ in this paper. The constructions for the $\bm{A_2}$ and $\bm{A_3}$ are very similar to that for this case. 

We have four zeros in the mass matrix $M^{0\,(A_1)}_d$ which is
\begin{align}
M^{0\,(A_1)}_d=
\begin{pmatrix}
0 & a & 0 \\
a' & 0& c\\
a'' & 0 & d
\end{pmatrix} \ , \quad
\label{Md10}
\end{align}
where $a, a',a'',c$ and $d$ are real parameters,  since we assume the CP invariance at the fundamental level. All Dirac type mass matrices $M$ in this paper is written as $M_{LR}$ form in \cite{Tanimoto:2016rqy}. 

We now generate spontaneous breaking of the CP invariance introducing two complex scalar bosons $\eta$ and $\eta'$ whose $\bm{Z_2}$ charges are odd($\bm{-}$). They are assumed to have complex vevs  at an intermediate scale  $\sim 10^{12}$ GeV. Note that the presence of two vevs, $<\eta>$ and $<\eta'>$ is necessary to break both of CP and the flavor $\bm{Z_2}$ symmetries \footnote{The spontaneous breaking produces domain walls in the early universe. However, they are diluted by the inflation.}. (See \cite{Liang:2024wbb} for details about the breaking mechanism.) 

We have two ways to localize the $\eta$ and $\eta'$ on the fixed points to obtain two zeros textures. One is to localize both on the fixed point I, and the other is to localize one on the fixed point I\hskip -0.5mm I and the other on the fixed point I\hskip -0.5mm I\hskip -0.5mm I. But the former does not work and hence  a successful mass matrix for the down-type quarks is given by, 
\begin{align}
M^{(A_1)}_d=
\begin{pmatrix}
0 & a & 0 \\
a' & \kappa<\eta>& c\\
a'' & \kappa'<\eta'> & d
\end{pmatrix} \ , \quad
\label{Md1}
\end{align}
where the $\eta$ and $\eta'$ terms are induced by heavy Higgs,
$\bm{H_{ I}}$ and $\bm{H_{I\hskip -0.5mm I}}$ exchanges, respectively, as shown in \cite{Liang:2024wbb}. The $\kappa = (f/M^2_{I})<H^\dagger>$ and $\kappa'= (f'/M^2_{II})<H^\dagger>$. Here $f$ and $f'$ are real dimension-one effective coupling constants, and $M_{I}$ and $M_{II}$ are the masses of the heavy Higgs bosons, $\bm{H_I}$ and $\bm{H_{II}}$.
The $<\eta> = B_\eta e^{i\phi}$ and $<\eta'> = C'_{\eta'} e^{i\phi'}$, where the $B_\eta$ and $C'_{\eta'}$ are real dimension-one constant parameters and $\phi$ and $\phi'$ are independent CP violating phases.

We rewrite the texture of the down-type quark mass matrix as 
\begin{align}
M^{\,(A_1)}_d=
\begin{pmatrix}
0 & a & 0 \\
a' & b e^{i\phi}& c\\
a'' & c'e^{i\phi'} & d
\end{pmatrix} \ ,\quad
\label{MdA1}
\end{align}
where $b=\kappa B_\eta$ and $c'=\kappa' C'_{\eta'}$, and  all parameters are real. We see that the determinant of this matrix is indeed real\,!

We see the textures of the down-type quark mass matrices for the type $A_2$ and $A_3$ as :

\begin{align}
M^{\,(A_2)}_d=
\begin{pmatrix}
0 &  0 &a \\
a' &c& b e^{i\phi}\\
a'' & d&c'e^{i\phi'} 
\end{pmatrix} \ ,\quad
\label{MdA2}
\end{align}
and

\begin{align}
M^{\,(A_3)}_d=
\begin{pmatrix}
 a & 0 & 0 \\
 b e^{i\phi} &a' & c\\
 c'e^{i\phi'}&a''  & d
\end{pmatrix} \ .\quad
\label{MdA3}
\end{align}

The matrices $M_d^{(A_2)}$ and $M_d^{(A_3)}$ are given by the exchanges of the second and third columns, and the first and second columns of $M_d^{(A_1)}$, respectively. Thus, we emphasize here that
the CKM matrix and mass eigenvalues are the same 
among $M_d^{(A_1)}$, $M_d^{(A_2)}$ and $M_d^{(A_3)}$. Hence, all parameters in those matrices obtained from the experimental data are the same.
We present, in Table \ref{tab:parameters}, the numerical values for each parameters derived from the experimental data on the quark masses and the CKM matrix elements
\cite{ParticleDataGroup:2024cfk},
where  we have taken the  2$\sigma$ error bars for the experimental data.

\begin{table}[hbtp]
\begin{center}
\begin{tabular}{|c|c|c|c|c|c|c|}
\hline 
	\rule[14pt]{0pt}{3pt}  
 $a/d \times 10^{2}$ & $a'/d\times 10^{2}$ & $|a''/d|\times 10^{2}$   & $b/d\times 10^{2}$  &  $c/d\times 10^{2} $ & $c'/d$ &$\phi'-\phi$ $^{[\circ]}$\\
\hline 
	\rule[14pt]{0pt}{3pt}  
 $0.70 \rightarrow 0.79$ &$0.40 \rightarrow 0.99$&$0\rightarrow 10$ & $4.3\rightarrow 4.9$ 
 &$3.6 \rightarrow 3.9$  & $0.79 \rightarrow 1.0$ &$37 \rightarrow 48$\\
\hline
\end{tabular}
\caption{The allowed range of parameters in  $M_d^{(A_1)}$,
$M_d^{(A_2)}$ and $M_d^{(A_3)}$,
where $2\,\sigma$ error bars of the quark masses and the CKM angles and  CP phase are taken.}
\label{tab:parameters}
\end{center}
\end{table}
The number of parameters in the up- and down-type quark mass matrices is twelve, but the observable parameters are  the  six quark masses and the four CKM matrix elements.
Therefore, two parameters must be redundant. This is the reason why only the difference between
the phases $\phi$ and $\phi'$ is constrained  as seen in Table \ref{tab:parameters}. In addition, it is also the reason why the parameter $a''$ is widely determined compared with other parameters.
We have taken positive values  for  $a$, $a'$, $b$, $c$, $c'$ and $d$ without loss of generality, but we should take both negative and positive  values for $a''$ in our numerical calculations. This convention for the parameters is possible because of the presence of two zeros
in the down-type quark mass matrix and the diagonal up-type quark one.

\subsection{The charged lepton mass matrices}
Let us discuss the charged lepton mass matrices in this subsection. The matrices are given by transposed matrices of the down-type quark mass matrices in 
Eqs.\,\eqref{MdA1}, \eqref{MdA2} and \eqref{MdA3} if the heavy Higgs bosons $\bm{H_I}$ and $\bm{H_{II}}$ belong to $\bm{5}$ of the $SU(5)_{GUT}$. However, those matrices can not reproduce the observed mass ratios, $m_e/m_{\mu}$ and $m_{\mu}/m_{\tau}$. Therefore, we consider the heavy Higgs bosons are mixtures of $\bm{5^*}$ and $\bm{45}$ of the $SU(5)_{GUT}$. Thus, we have two free parameters $k_e$ and $k'_e$ \cite{Tanimoto:2024nwm}
which represent the two mixing parameters between $\bm{5^*}$ and $\bm{45}$.

Now, the textures of the charged lepton mass matrices are given for each type, $\bm{A_1}$, $\bm{A_2}$ and $\bm{A_3}$ as follows;

\begin{align}
M^{\,(A_1)}_e=
\begin{pmatrix}
0 & a' & a''\\
a & k_e b e^{i\phi}& k_e' c'e^{i\phi'}\\
0 & c & d
\end{pmatrix} \ ,\quad
\label{MeA1}
\end{align}

\begin{align}
M^{\,(A_2)}_e=
\begin{pmatrix}
0 & a' & a''\\
0 & c & d\\
a & k_e b e^{i\phi}& k_e' c'e^{i\phi'}
\end{pmatrix} \ ,\quad
\label{MeA2}
\end{align}

\begin{align}
M^{\,(A_3)}_e=
\begin{pmatrix}
a & k_e b e^{i\phi}& k_e' c'e^{i\phi'}\\
0 & a' & a''\\
0 & c & d
\end{pmatrix} \ .\quad
\label{MeA3}
\end{align}

\begin{figure}[H]
\begin{minipage}[]{0.47\linewidth}
\includegraphics[width=8cm]{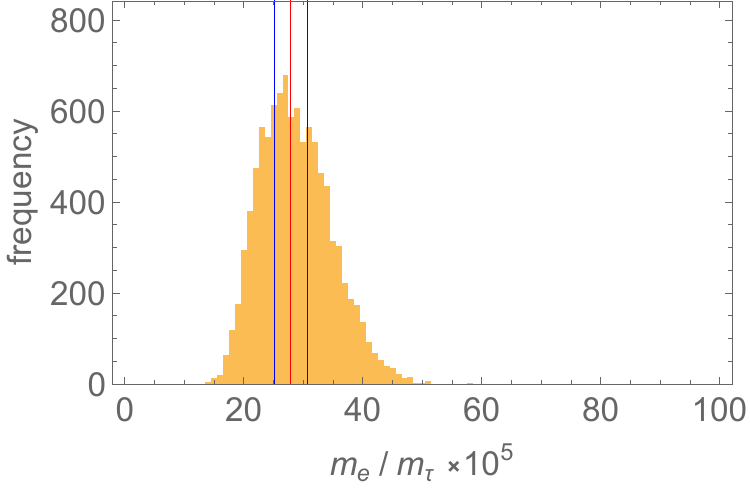}
\caption{ The predicted distribution of $m_e/m_{\tau}$ by taking $k_e=3$ and $k_e'=1$ in the case of $\bm{A_1}$. The vertical red line denotes the central value of the observed one, and blue ones denote $\pm 10\%$
error-bars for eye guide.}
\label{fig:mass13}
\end{minipage}
\hspace{2mm}
\begin{minipage}[]{0.47\linewidth}
\vskip -0.5 cm
\includegraphics[width=8cm]{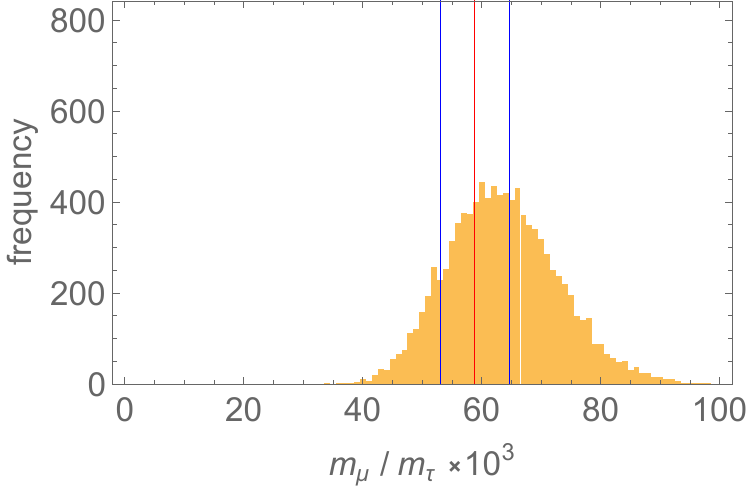}
\caption{The predicted distribution of $m_{\mu}/m_{\tau}$ by taking $k_e=3$ and $k_e'=1$ in the case of $\bm{A_1}$. The vertical red line denotes the central value of the observed one, and blue ones denote $\pm 10\%$
error-bars for eye guide.}
\label{fig:mass23}
\end{minipage}
\end{figure}

We reproduce the observed  mass ratios $m_e/m_{\mu}$ and $m_{\mu}/m_{\tau}$ at the electroweak scale \cite{Antusch:2013jca} by taking $k_e=3$ and $k_e'=1$
 \footnote{The mass ratios of the charged leptons are almost
 unchanged by the evolution of the renormalization group equations.}. 
 The results on mass ratios $m_e/m_{\mu}$ and $m_{\mu}/m_{\tau}$ are presented, respectively in Figs.\,\ref{fig:mass13} and \ref{fig:mass23}
\footnote{Frequency plots in Figs.\,\ref{fig:mass13} and \ref{fig:mass23} are obtained by assuming the normal distribution  of parameters in Table \ref{tab:parameters}.
 The standard deviations are taken to be 2$\sigma$ for the listed parameter ranges.}.
Notice that the mass 
 eigenvalues are all the same among in the three cases, $\bm{A_1, \,A_2, \,A_3}$. Thus, the mass ratios are the same, too.
However, the mixing matrices of the left-handed charged leptons are all different for each others in the three cases $\bm{A_1,A_2,A_3}$. This becomes very important when we consider the neutrino oscillation in the next section.

\section{The neutrino mass matrices}
We introduce three right-handed neutrinos $N_{1}$,$N_{2}$ and $N_{3}$, and put them on each fixed point I,\,I\hskip -0.5mm I,\,I\hskip -0.5mm I\hskip -0.5mm I, separately (see Table 1). Their $Z_2$ 
charges are the same as $\bm{10}$'s.
Therefore, $N_{1,2,3}$ have 
$(\bm{-},\bm{+},\bm{+})$.
The heavy Majorana mass matrix is real diagonal as the up-type quark mass matrix. The Dirac mass matrices for the neutrinos, $M_D$, are given by the same forms of the texture of the charged lepton mass matrices 
in Eqs.\,\eqref{MeA1}, \eqref{MeA2} and \eqref{MeA3}, but the parameters of the matrix elements are different except for the phases $\phi$ and $\phi'$ 
\cite{Tanimoto:2024nwm}.

The neutrino mass matrix is given by the seesaw mass formula $M_\nu \simeq M_D (1/M_N) M_D^t$ \cite{Minkowski:1977sc,Yanagida:1979as,Yanagida:1979gs,Gell-Mann:1979vob}, where $M_N$ is the diagonal right-handed Majorana mass matrix, and hence we have additional unknown parameters, that is, $M_1, M_2, M_3$. However, we absorb the $1/\sqrt{M_i}$ in the Dirac mass matrix $M_D$ and hence the number of free parameters in the neutrino mass matrix $M_\nu$ is nine. The redefined Dirac mass matrices for the each types are written as
\begin{align}
M'^{\,(A_1)}_D=
\begin{pmatrix}
0 & A' & A'' \\
A & B\, e^{-i\phi} & C'\, e^{-i\phi'} \\
0 & C& D
\end{pmatrix}\ , \quad
\label{MDA1}
\end{align}

\begin{align}
M'^{\,(A_2)}_D=
\begin{pmatrix}
0 & A' & A'' \\
0 & C& D\\
A & B\, e^{-i\phi} & C'\, e^{-i\phi'}
\end{pmatrix}\ , \quad
\label{MDA2}
\end{align}

\begin{align}
M'^{\,(A_3)}_D=
\begin{pmatrix}
A & B\, e^{-i\phi} & C'\, e^{-i\phi'} \\
0 & A' & A'' \\
0 & C& D
\end{pmatrix}\ . \quad
\label{MDA3}
\end{align}
And the neutrino mass matrix $ M_\nu $ is written as 
\begin{equation}
    M_\nu \simeq M'_D M_D^{'t}\,.
\end{equation}

\begin{figure}[H]
\begin{minipage}[]{0.47\linewidth}
\includegraphics[width=8cm]{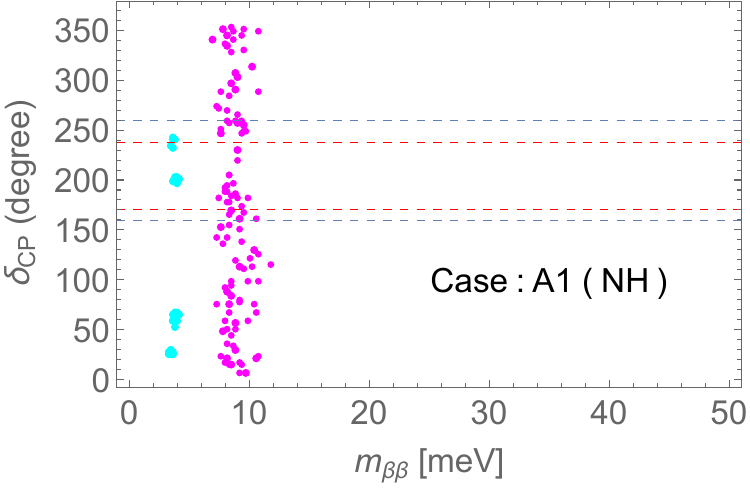}
\caption{The predicted $\delta_{CP}$ versus  $m_{\beta\beta}$
for NH in the case of  $\bm{A_1}$.
The region between the  horizontal red (blue) dashed-lines 
denotes $1\,(2)\sigma$ allowed one of $\delta_{CP}$ in NuFIT 6.0  (NH with SK atmospheric data) \cite{Esteban:2020cvm}.
The cyan and magenta regions denote the regions of $m_1\ll m_2$ and $m_1\lesssim m_2$, respectively.}
\label{fig:A1NH-mee-CP}
\end{minipage}
\hspace{2mm}
\begin{minipage}[]{0.47\linewidth}
\vskip -1 cm
\includegraphics[width=8cm]{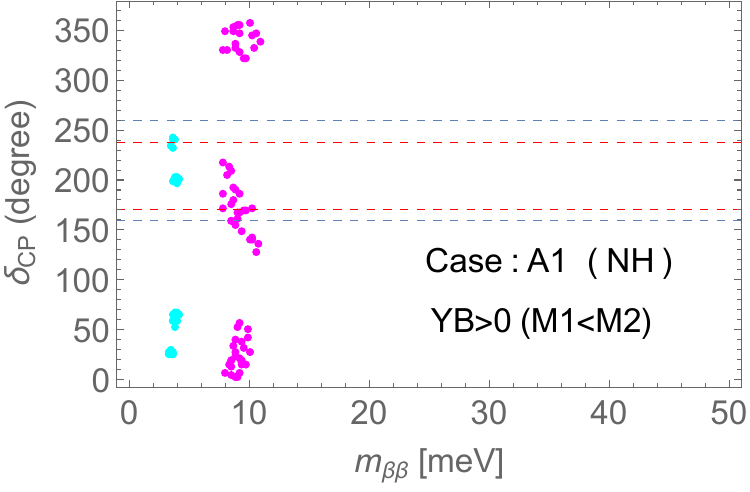}
\caption{The predicted $\delta_{CP}$ versus  $m_{\beta\beta}$
for NH in the case of $\bm{A_1}$ by putting the constraint the positive cosmological baryon number $Y_B>0$ for the case of  $M_1< M_2$.
The notations are same as in Fig.\,\ref{fig:A1NH-mee-CP}.
}
\label{fig:A1NH-YB+mee-CP}
\end{minipage}
\end{figure}

We find that all cases of  $\bf{A_{1}}$, $\bf{A_{2}}$ and $\bf{A_{3}}$ predict the normal hierarchy (NH) in the neutrino mass eigenvalues $m_1$,\,$m_2$
and $m_3$. The inverted hierarchy (IH) is very difficult to realize for all cases. We show, in Fig.\,\ref{fig:A1NH-mee-CP}, predictions on the CP violating parameter $\delta_{CP}$ in the neutrino oscillation and the $m_{\beta\beta}$ for the double $\beta$ decays of nuclei.
In addition, we  impose the positive cosmological baryon asymmetry,
$Y_B>0$ \cite{Tanimoto:2024nwm} assuming the first Majorana right-handed neutrino $N_1$ is the lightest among the heavy right-handed neutrinos,  $M_1< M_{2},\,M_{3}$.  The result is presented
in Fig.\,\ref{fig:A1NH-YB+mee-CP}.
{Some regions in Fig.\,\ref{fig:A1NH-mee-CP}  are excluded due to $Y_B>0$.}

We see that the CP violating phase $\delta_{CP}$ is predicted in a very broad region, but the $m_{\beta\beta}$ is predicted in two narrow regions, $m_{\beta\beta}\simeq 4$meV and  $(8\rightarrow 10)$meV.
{The total sum of neutrino masses 
$\sum m_i$ is given as $58$meV$\rightarrow 60$meV and
 $65$meV$\rightarrow 75$meV,
which are consistent with the present experimental bound 
(see Appendix A).}

It is interesting that we can explain the correct sign of the baryon asymmetry in the present universe for $M_1<M_2$ in the region where the predicted $\delta_{CP}$ is consistent with the observation within the one standard deviation $1\,\sigma$ \footnote{See \cite{Tanimoto:2024nwm} for the calculation of the sign of the baryon asymmetry.}.  We find all results in the three cases $\bf{A_{1}}$, $\bf{A_{2}}$ and $\bf{A_{3}}$ are almost the same.

\section{The case of two right-handed neutrinos}

The heavy Majorana neutrinos can produce the baryon asymmetry in the universe through the leptogenesis \cite{Fukugita:1986hr}. However, two heavy Majorana neutrinos are enough to explain the observed bayon asymmetry in the present universe \cite{Frampton:2002qc}. In this section we introduce two right-handed neutrinos $N_i ~(i=1,2)$ or $(i=2,3)$. In practice we take one of three right-handed neutrinos superheavy by taking its mass infinity. We consider two cases, that is,  $M_1 = \infty$ and $M_3 =\infty$ in the previous section \footnote{We have found the almost same predictions in the case of $M_2 = \infty$ as in the case of $M_3 =\infty$.}. Then,   all elements of the  first  and third column
 of the Dirac mass matrices of 
 Eqs.\,\eqref{MDA1}, \eqref{MDA2} and \eqref{MDA3}
 vanish for the case of $M_1 = \infty$ and $M_3 = \infty$, respectively.
 Since the number of parameters are reduced
  due to  $M_1 = \infty$ and $M_3 = \infty$, the CP violating phases $\delta_{CP}$ are predicted in more narrow regions. 

\begin{figure}[H]
\begin{minipage}[]{0.47\linewidth}
\includegraphics[width=8cm]{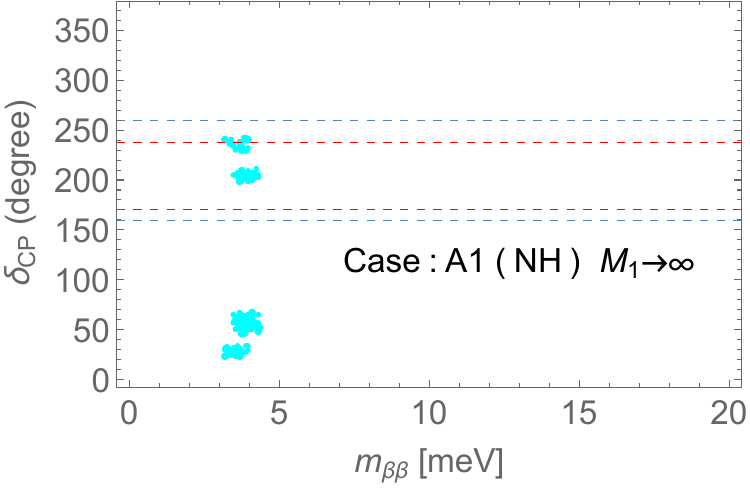}
\caption{The predicted $\delta_{CP}$ versus  $m_{\beta\beta}$
for NH in the case of $A_1$ with infinite  $M_1$.
The region between the  horizontal red (blue) dashed-lines 
denotes $1\,(2)\sigma$ allowed one of $\delta_{CP}$ in NuFIT 6.0  
\cite{Esteban:2020cvm}. }
\label{fig:A1NH-M1infty}
\end{minipage}
\hspace{2mm}
\begin{minipage}[]{0.47\linewidth}
\vskip -0.5 cm
\includegraphics[width=8cm]{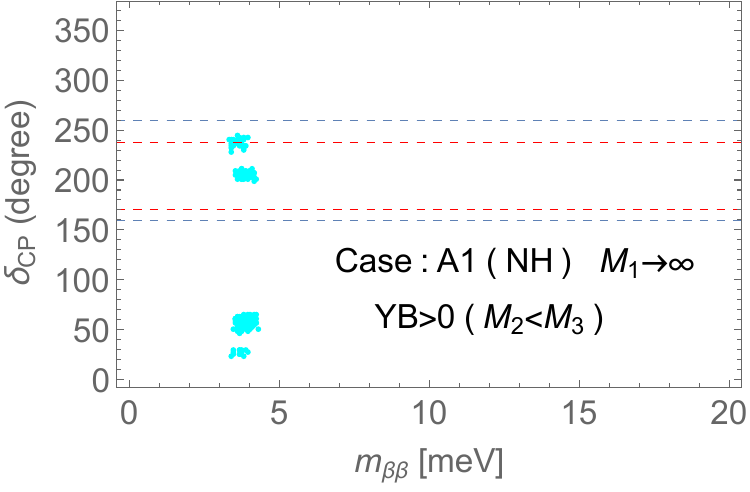}
\caption{The predicted $\delta_{CP}$ versus  $m_{\beta\beta}$
for NH in the case of $A_1$ with infinite  $M_1$ by putting the constraint the positive cosmological baryon number $Y_B>0$ for the case of $M_2< M_3$.
}
\label{fig:A1NH-M1infty-YB+}
\end{minipage}
\end{figure}

In Fig.\,\ref{fig:A1NH-M1infty}, we show the prediction of  $\delta_{CP}$ and $m_{\beta\beta}$ for the case of  $M_1 = \infty$.
We, furthermore, impose the positive cosmological baryon asymmetry,
$Y_B>0$ for the case of  $M_2< M_3$.
In Fig.\,\ref{fig:A1NH-M1infty-YB+}, we show its prediction of  $\delta_{CP}$ and $m_{\beta\beta}$,
{which are almost same as the ones in Fig.\,\ref{fig:A1NH-M1infty}.} 
It is interesting that the CP violating phase $\delta_{CP}$ is predicted in narrow regions, one of which is consistent with  the experimental data.
The total sum of neutrino masses 
$\sum m_i$ is given as $58$meV$\rightarrow 59$meV, and $m_{\beta\beta}\simeq 4$meV.
We should comment here that the almost same results are obtained in the other two cases, $\bm{A_2, A_3}$.

\begin{figure}[H]
\begin{minipage}[]{0.47\linewidth}
\includegraphics[width=8cm]{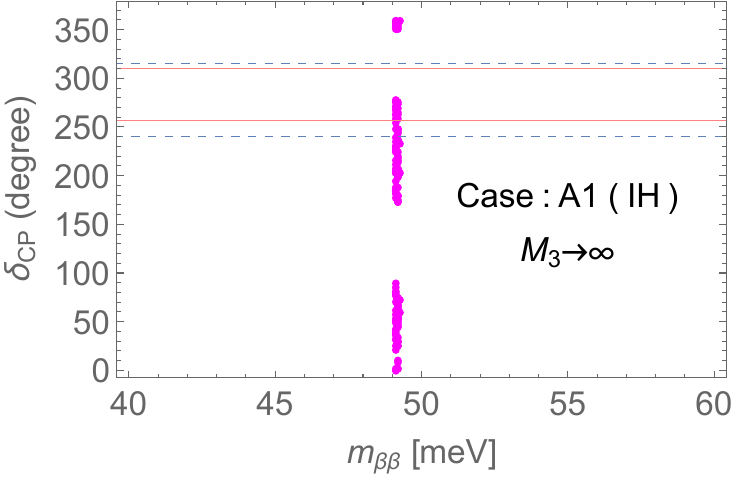}
\caption{The predicted $\delta_{CP}$ versus  $m_{\beta\beta}$
for IH in the case of $A_1$ with infinite  $M_3$.
The region between the  horizontal red (blue) dashed-lines 
denotes $1\,(2)\sigma$ allowed one of $\delta_{CP}$ in NuFIT 6.0  (IH without SK atmospheric data) \cite{Esteban:2020cvm}. }
\label{fig:A1IH-M3infty}
\end{minipage}
\hspace{2mm}
\begin{minipage}[]{0.47\linewidth}
\vskip -0.5 cm
\includegraphics[width=8cm]{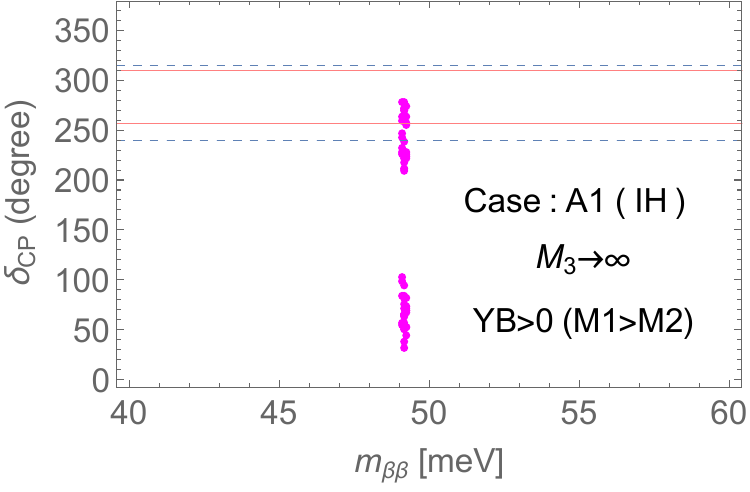}
\caption{The predicted $\delta_{CP}$ versus  $m_{\beta\beta}$
for IH in the case of $A_1$ with infinite  $M_3$ by putting the constraint the positive cosmological baryon number $Y_B>0$ for the case of  $M_1> M_2$.
}
\label{fig:A1IH-M3infty-YB+}
\end{minipage}
\end{figure}
As for the case of $M_3 = \infty$, we find that only the inverted hierarchy (IH) is realized. 
In Fig.\,\ref{fig:A1IH-M3infty}, we show the prediction of  $\delta_{CP}$ and $m_{\beta\beta}$ for   $M_3 = \infty$. We see the broad region for the $\delta_{CP}$ while  $m_{\beta\beta}\simeq 49$meV.
{The total sum of neutrino masses 
$\sum m_i$ is given as $99$meV,
which is still consistent with the present experimental bound 
(see Appendix A).}
In Fig.\,\ref{fig:A1IH-M3infty-YB+}, we show also the predictions of  $\delta_{CP}$ and $m_{\beta\beta}$ imposing the positive baryon asymmetry in the universe. 
{It is found that  some regions 
in Fig.\,\ref{fig:A1IH-M3infty} are excluded.}
From both figures, we see the positive cosmological baryon asymmetry, $Y_B>0$, in the present universe implies that the first family right-handed neutrino is heavier than the second one, $M_1> M_2$.

We also comment that the almost same results are obtained in the other two cases, $\bm{A_2, A_3}$
even in the case of two right-handed neutrinos.

\section{Conclusions}

We have constructed two zeros textures of the down-type quark mass matrices based on the $\bm{T^2/Z_3}$ orbifold  in the six dimensional space-time. We have realized a diagonal mass matrix for the up-quark mass matrices, taking the size $L$ of compactification a bit larger than the fundamental cut-off scale $M_*$ ($LM_* > 30$ \cite{Liang:2024wbb}). Here, we have put all $\bm{10_i}$ on the separate three fixed points in the extra two dimensions. All $\bm{5^*_i}$ are living in the bulk. To generate zeros in mass matrices for the down-type quarks, we have imposed a flavor discrete symmetry $\bm{Z_2}$.  Here, all of them have four zeros in the textures. All elements are real, since we have assumed the CP invariance at the fundamental level. 

We have introduced two vevs of the complex scalar bosons $\eta$ and $\eta'$ to break the CP- and $\bm{Z_2}$ symmetries, which makes two zeros textures of the down-type quark mass matrices. Then, we have found 9 type textures for the down-type quark mass matrices whose determinants are real. This provides us with a solution to  the strong CP problem. However, we have found only three type textures remain consistent with the CKM observations, which we have called  $\bm{A_1}$, $\bm{A_2}$ and $\bm{A_3}$ types. We have derived all real parameters including the difference between the two phases in the matrix from the experimental data of the CKM matrix and the observed quark masses.

We have extended the above model including the three right-handed neutrinos $N_i ~(i=1,2,3)$ to generate the neutrino masses and mixing. We have put them on the each fixed points as for the case of
$\bm{10_i ~(i=1,2,3)}$. We have found this model predicts naturally the normal hierarchy (NH) of the neutrino masses. The three type textures $\bm{A_1}$, $\bm{A_2}$ and $\bm{A_3}$ predict the broad range of the CP violation parameter, $\delta_{CP}$, as shown in Fig.\,\ref{fig:A1NH-mee-CP}. On the contrary, the key parameter, $m_{\beta\beta}$, for the double $\beta$ decays of nuclei is predicted in a narrow region $m_{\beta\beta} \simeq 4$meV and $(8\rightarrow 10)$meV as shown in Fig.\,\ref{fig:A1NH-mee-CP}. 
{It is remarkable that all of the three type textures $\bm{A_1}$, $\bm{A_2}$ and $\bm{A_3}$ have almost the same predictions on $\delta_{CP}$ and  $m_{\beta\beta}$. As expected the prediction on the $\delta_{CP}$ is much  broader than that in the previous three zero texture and hence we can distinguish both case in the future experiments. As for the $m_{\beta\beta}$ if it is smaller than $\sim 8$meV the previous three zero texture model will be excluded.}

We have also considered the case of two right-handed neutrinos $N$s, since it is known that two right-handed neutrinos are enough to produce the lepton asymmetry in the early universe (leptogenesis) \cite{Frampton:2002qc}. We have taken $M_1=\infty$ or $M_3=\infty$ to represent effectively the two cases of the two right-handed neutrinos.

We have found the normal hierarchy (NH) for the first case $M_1=\infty$ and the inverted hierarchy (IH) for the second case $M_3=\infty$. We should stress here that the IH is realized only in the two zero texture models. Here, the presence of two phases is crucial to realize the IH.
We have found the NH case has a very narrow prediction  on $\delta_{CP}$ and on $m_{\beta\beta}$ as shown in Figs.\,\ref{fig:A1NH-M1infty}
and \ref{fig:A1NH-M1infty-YB+},
$m_{\beta\beta}\simeq 4$meV and 
 $\delta_{CP} \simeq 200^\circ$ or $250^\circ$. 
It is very remarkable that the positive cosmological baryon asymmetry 
is reproduced for $M_2< M_3$ in this NH case \footnote{The missing right-handed neutrino can be the dark matter, where we consider its Yukawa couplings are vanishing instead of taking its mass infinity $M_i=\infty$.}.

\section*{Acknowledgments}
T.~T.~Y.~thanks Qiuyue Liang for discussion about two-zeros texture of quark and lepton mass matrices. This work was in part supported by MEXT KAKENHI Grants No.~24H02244 (T.~T.~Y.). T.~T.~Y.~was supported also by the Natural Science Foundation of China (NSFC) under Grant No.~12175134 as well as by World Premier International Research Center Initiative (WPI Initiative), MEXT, Japan. 

 \appendix
\section*{Appendix}
\label{Data}
\section{Input Data}
We input the data for the case of NH,
\begin{table}[H]
	\begin{center}
		\begin{tabular}{|c|c|c|}
			\hline
			      \rule[14pt]{0pt}{2pt}
\ observable\       &                 best fit\,$ \pm 1\,\sigma$ for NH                  &          2 $\sigma $  range for NH  \\ \hline
			          \rule[14pt]{0pt}{2pt}				                                          $\sin^2\theta_{12}$               &                     $0.308^{+0.012}_{-0.011}$                     &       $0.28\rightarrow 0.33$                \\                 \rule[14pt]{0pt}{2pt}
			$\sin^2\theta_{23}$   &                     $0.470^{+0.017}_{-0.013}$                     &               $0.42\rightarrow 0.59$                \\
			          \rule[14pt]{0pt}{2pt}
			              $\sin^2\theta_{13}$               &                  $0.02215^{+0.00056}_{-0.00058}$            
			              &    $0.021\rightarrow 0.023$    \\ 
			          \rule[14pt]{0pt}{2pt}
$\Delta m_{21}^2$  & $7.49^{+0.19}_{-0.19}\times 10^{-5}{\rm eV}^2$  &     $(7.11 \rightarrow 7.87)\times  10^{-5}\,{\rm eV}^2$     \\ 
\rule[14pt]{0pt}{2pt}
$\Delta m_{31}^2$ & \ \ \ \ $2.513^{+0.021}_{-0.019}\times 10^{-3}{\rm eV}^2$ \ \ \ \ & $(2.47 \rightarrow  2.56)\times 10^{-3}\,{\rm eV}^2$ \ \ \\  \hline
		\end{tabular}
\caption{The best fit\,$ \pm 1\,\sigma$ and $2\sigma$ range of neutrino  parameters from NuFIT 6.0 for NH (with SK atmospheric data)
		\cite{Esteban:2020cvm}.
		}
		\label{DataNufit-NH}
	\end{center}
\end{table}
and for the case of IH,
\begin{table}[H]
	\begin{center}
		\begin{tabular}{|c|c|c|}
			\hline
			      \rule[14pt]{0pt}{2pt}
\ observable\       &                 best fit\,$ \pm 1\,\sigma$ for IH                  &          2 $\sigma $  range for IH  \\ \hline
			          \rule[14pt]{0pt}{2pt}				                                          $\sin^2\theta_{12}$               &                     $0.308^{+0.012}_{-0.011}$                     &       $0.28 \rightarrow 0.33$                \\                 \rule[14pt]{0pt}{2pt}
			$\sin^2\theta_{23}$   &                     $0.562^{+0.012}_{-0.015}$                     &               $0.45 \rightarrow 0.59$                \\
			          \rule[14pt]{0pt}{2pt}
			              $\sin^2\theta_{13}$               &                  $0.02224^{+0.00056}_{-0.00057}$            
			              &    $0.021 \rightarrow  0.023$    \\ 
			          \rule[14pt]{0pt}{2pt}
$\Delta m_{21}^2$  & $7.49^{+0.19}_{-0.19}\times 10^{-5}{\rm eV}^2$  &     $(7.11 \rightarrow 7.87)\times  10^{-5}\,{\rm eV}^2$     \\ 
\rule[14pt]{0pt}{2pt}
$\Delta m_{31}^2$ & \ \ \ \ $-2.510^{+0.024}_{-0.025}\times 10^{-3}{\rm eV}^2$ \ \ \ \ & $-(2.46 \rightarrow 2.56)\times 10^{-3}\,{\rm eV}^2$ \ \ \\  \hline
		\end{tabular}
\caption{The best fit\,$ \pm 1\,\sigma$ and $2\sigma$ range of neutrino  parameters from NuFIT 6.0 for IH (without SK atmospheric data)
		\cite{Esteban:2020cvm}.
		}
		\label{DataNufit-IH}
	\end{center}
\end{table}

 In addition, one should take into account the constraints from the CP Dirac phase
 $\delta_{CP}$ of NuFIT 6.0  \cite{Esteban:2020cvm}  and 
 the cosmological bound of the total sum of  neutrino masses \cite{Vagnozzi:2017ovm,Planck:2018vyg,Gerbino:2022nvz}.
{The CP  Dirac phase is given by at the $1\sigma$ level:
\begin{equation}
  \delta_{CP} = (\,212^{+26}_{-41}\,)^{\, \circ}\quad {\rm for\ \ NH\ \ and}
    \quad (\,285^{+25}_{-28}\,)^{\, \circ}\quad {\rm for\ \ IH}\,,
  \label{CP}
\end{equation}
}
and  the total sum of neutrino masses is given as:
\begin{equation}
  \sum_{i=1}^3 m_i<120\ {\rm meV}\,.
   \label{sum}
\end{equation}


\bibliographystyle{JHEP} 

\bibliography{ref}

\end{document}